\documentclass[runningheads]{llncs}
\usepackage{graphicx}
\usepackage{wrapfig}
\usepackage{float}
\usepackage{amsmath}
\usepackage{amssymb}
\usepackage{mathtools}
\usepackage[english]{babel}
\usepackage{cite}
\usepackage{textcmds}
\usepackage[utf8]{inputenc}
\usepackage{lineno}
\usepackage{dirtytalk}
\usepackage{xspace}
\usepackage{wrapfig}

\newcommand{\Heraklit}{\normalfont \textsc{Heraklit}\xspace}

\begin{document}
\title{Towards a fundamental theory of modeling discrete systems}
\titlerunning{Towards a fundamental theory}
%
%

\author{Peter Fettke\inst{1,2}\orcidID{0000-0002-0624-4431} \and
Wolfgang Reisig\inst{3}\orcidID{0000-0002-7026-2810}}
\authorrunning{P. Fettke, W. Reisig}
%

\institute{German Research Center for Artificial Intelligence (DFKI), Saarbr\"ucken, Germany \\
\email{peter.fettke@dfki.de}\\ \and
Saarland University, Saarbr\"ucken, Germany \\ \and
Humboldt-Universität zu Berlin, Berlin, Germany \\ 
\email{reisig@informatik.hu-berlin.de}}

\maketitle 
\begin{abstract}
Modeling is a central concern in both science and engineering. However, we need a new fundamental theory to address the challenges of the digital age. In this paper, we first explain why modeling is fundamental and which challenges must be addressed in the digital world. As a main contribution, we introduce the \Heraklit modeling framework as a new approach to modeling. We conclude with some general remarks. Future work will involve the correctness of modeling, the notion of information, and the description of invariance in modeling.

\end{abstract}

\section{Part I: Modeling computer-integrated systems}

\subsection{Modeling as a central concern of science and engineering}
The use of models is a central concern of science. As John von Neumann emphasizes in \cite{Neumann55}: 

\begin{quote}
The sciences do not try to explain, they hardly even try to interpret, they mainly make models. By a model is meant a mathematical construct which, with the addition of certain verbal interpretations, describes observed phenomena. The justification of such a mathematical construct is solely and precisely that it is expected to work—that is, correctly to describe phenomena from a reasonably wide area.
\end{quote}

This quote rises the quest for adequate modeling framework. Generally stated, a good formal modeling technique allows a modeler to express his thoughts most faithfully and unambiguously in a formal framework. Bound to concrete examples, we have usually a clear understanding of what an \say{adequate} description of a real-world or imagined system could be: it should cover all aspects one would like to emphasize and it should hide all aspects one would prefer not to mention.

As a way to gain faithful models, the axiomatic method is particularly useful. One starts with a set of basic assumptions (axioms) and then derives further statements (theorems) through logical deduction. The set of axioms should be as small as possible, and each axiom should be carefully motivated and justified.

For models of discrete systems, three aspects are crucial: The identification of components of the system, the level of abstraction of items, and the description of dynamics. A model provides means to better understand the world. Reality either is given or, as in engineering, to be constructed by experts. The century long efforts, far from completion, of the physics community to achieve a unifying theory as a basis for all branches of physics provides an impressive paradigm of substantial scientific progress. 

This rises a number of questions about fundamental problems for a theory of discrete systems modeling: Is this kind of faithful modeling also conceivable for discrete models of the digital word? Are there some general principles to construct such models? What role do conceptual models play in computing?

\subsection{From algorithms to digital systems}

The digital world includes classical mathematical algorithms, as well as numerous other systems that incorporate computers. Typical examples include:

\begin{itemize}

\item Euklid’s algorithm, computing the greatest common divisor of two natural numbers;

\item the bisection algorithm, assuming as input three real numbers $a$, $b$, $\epsilon$, and a continuous function $f$ with $f(a) < 0$ and $f(b) > 0$. The algorithm returns a real number $a < c < b$ with ${\mid f(c) \mid} < \epsilon$; 

\item the alternating bit protocol, detecting and correcting loss of messages;

\item the hiring procedure of new staff in a company;

\item the organization of a bakery;

\item a robot, assembling a circuit board;

\item a cooking recipe; 

\item apple sort, with a given inclined board with holes of increasing size: apples roll down the board, eventually through a hole;

\item organization of a stock exchange.

\end{itemize}

Furthermore, the digital world includes big systems such as business information systems, embedded systems, the internet of things, cyber-physical systems, digital ecosystems, Industry 4.0, and digital infrastructures in the areas of health, mobility, industry, services, administration etc.

Which of the above systems are conceived as algorithms, is somewhat fruitless. Most of them are not intended to be implemented as a monolithic software. In order to crystallize a theory of such systems, it is useful to identify what they have in common. We identify three aspects: firstly, a \say{big} such system is usually composed of smaller systems (components, modules). Secondly, those systems deal with data as well as real world items. And thirdly, those systems proceed in discrete steps. These three aspects may serve as a starting point for a modeling theory of digital systems.

\subsection{Models in informatics}
The above quote of von Neumann rises the question, whether informatics is a scientific area at all: Which phenomena are observed and described? Informatics constructs software for many areas, such as business processes, technical production systems, and administrations. There is no need to observe or to describe this software. What must be observed or be described, i.e. what must be modeled, are the systems, viz. the business processes, technical production systems, and administrations, in which the software takes effect. Deep knowledge of the (intended) effect of a software in such a system, is a premise for the sensible construction of the software. Predicates for input and output of the software do not suffice. In this respect, informatics is conceptually weak. Generally accepted and approved such modeling frameworks are missing. And there is not much willingness to develop and apply such techniques. Existing frameworks such as ERM, UML, BPMN, and ARIS are by far not expressive enough to model all decisive aspects of digital systems.

In particular, the collective term “formal methods” covers mathematical methods to specify, analyze, and verify software. But also these methods are limited to pre- and postconditions and are not intended to model the role of software in real world systems.

\subsection{Three postulates on modeling computer-integrated systems}
A theoretical basis to bridge the gap between computation technology and its usage must meet various requirements. We formulate what we consider the central requirements, in terms of three postulates:

\begin{itemize}

\item Postulate 1: A comprehensive model of computer-integrated systems is structured. A large computer-integrated system is not amorphous, but is composed of a number of sub-systems. In one sentence: Composition matters!

\item Postulate 2: A model may include both data and real-life items: products, customers, machines, paper money, contracts, etc. In one sentence: Objects matter!

\item Postulate 3: A model proceeds locally. Behavior cannot be understood as sequence of steps in one, “global” state space. Instead, every change of the system has its local causes and local effects. Behavior evolves from local events. Some events are causally related; others are not. In one sentence: Causality matters!
\end{itemize}

Many ideas contributing to these postulates have been published before. But the time is ripe to integrate those hitherto only loosely related concepts. This is not achievable by just constructing a new “hyper model” of computation. Instead, fundamental quests on basic notions of computing are to be posed and answered, resulting in a basic framework for modeling discrete systems. In this perspective, the new model must conceive the traditional sequential calculation model not as an alternative, but as a particularly important, nevertheless special case.

\section{Part II: The modeling framework \Heraklit}
On the background of the above considerations, in particular the postulates described before, we suggest the modeling framework \Heraklit. It is motivated and justified by epistemological arguments. It combines well established and more recently suggested concepts.

\subsection{The epistemic basis of \Heraklit}
In the spirit of the axiomatic method as outlined before, \Heraklit assumes the absolute minimum of axioms: just the axioms of well-known elementary (Zermelo-Fraenkel) set theory with most elementary operations, such as union, intersection element of, tuples, functions, and predicates. This approach rules out any kind of implicit, hidden assumptions and unjustified arguments. Meyer and Weber in \cite{meyer2025programmingreallysimplemathematics} emphasize the advantages for this kind of starting a theory, and apply it to programming languages. 

We apply this method to the definition of \Heraklit. As a more convenient starting point, we replace sets by heterogeneous structures. This is no substantial difference, because a heterogeneous structure can be derived from sets by simple mathematics: a heterogeneous structure is just a finite collection of sets, distinguished elements (\say{constants}), functions, and predicates over these sets. Heterogeneous structures are most general: first order logic is interpreted in heterogeneous structures.  

Each heterogeneous structure can be attached a signature: each constant, function and predicate is assigned a symbol. Together with a set of variables, the constant- and function symbols generate (in general infinitely many) terms (symbol sequences) such as e.g. $f(g(a), x)$, where each interpretation of the variables yields a value in one of the structure’s sets. 

In many cases, one is not interested in just one, but in all structures with the same signature, or the structures with distinguished properties. All this is approved mathematical standard. Heterogeneous structures are explicitly or implicitly used in the context of various modeling frameworks.

\subsection{Modeling dynamics in discrete systems}
As outlined before, in informatics, dynamics evolves in discrete steps. This is usually modeled in terms of transition systems, i.e. graph structures, with each node representing a state, and each arc a step. A single behavior is then a path through the graph. Specification- and analysis techniques of temporal logic usually assume a heterogeneous structure, including in particular some predicates, and provides logical formulae that are interpreted in single behaviors or in sets of behaviors.

\Heraklit substantially differs from this approach, governed by a fundamental concept: dynamics is represented as updates of predicates. In a state $s$, a predicate $p$ may apply to an item $a$; a step then may cease this fact. Vice versa, in a state $s$, the predicate $p$ may not apply to an item $a$; a step then may cease this fact. In general, a \Heraklit step locally updates a bounded set of predicates. This concept generates an amazing theory of models for discrete systems. Next, we concentrate on the smooth step from static to dynamic predicates. We do so by means of an everyday example.

\subsection{Example: elementary steps in a restaurant organization}
We start to identify two predicates, \emph{waiting clients} and \emph{free tables}. Graphically, each predicate is represented as an ellipse in Fig. 1a.

\begin{figure}[h]
   \centering
   \includegraphics[scale=.45]{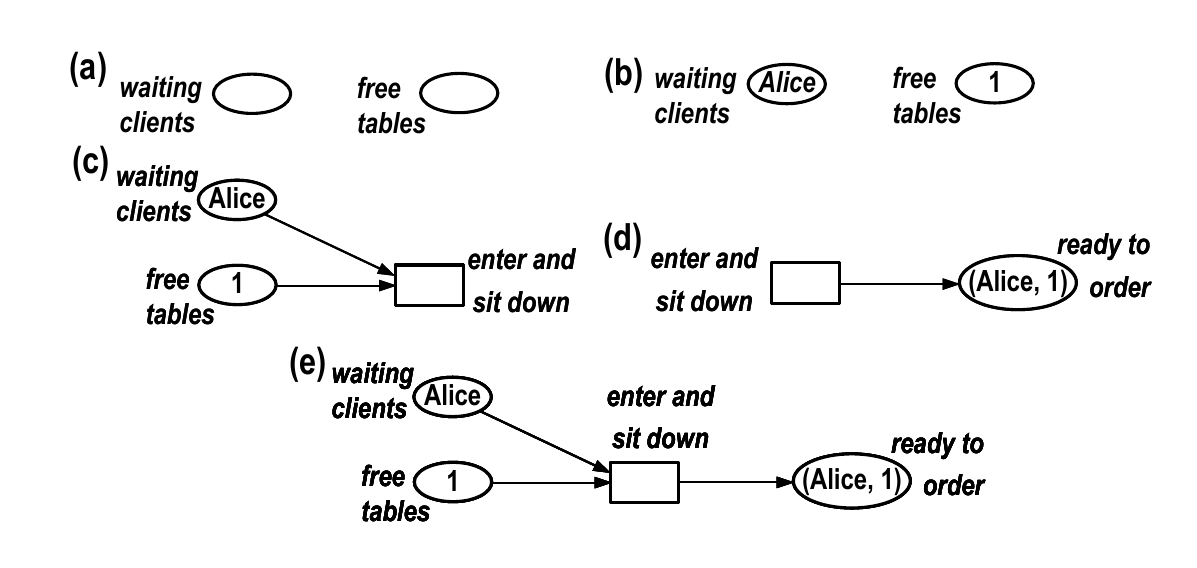}
   \caption{Elementary steps}
   \label{fig:01}
\end{figure}



To express the proposition that a client is waiting, its name is inscribed into the ellipse. Likewise, the number of a free table may be inscribed into the ellipse of the free tables predicate (Fig. 1b).



Predicates and propositions are nothing new: A predicate applies to an item, or does not apply to it. What’s now new is a systematic capturing of steps: Whether or not a predicate applies to an item, may change in an evolving system. 

In our running example, the so far waiting client \emph{Alice} may enter the restaurant and sit down at table 1. Then, the two propositions of Fig. 1b no longer hold. Graphically, this is expressed by means of a box, and arrows between the propositions and the box (Fig. 1c). Intuitively stated, the arrows \say{remove} the items \emph{Alice} and $1$ from the predicates. In an analogous manner, an item may \say{enter} a predicate (Fig. 1d). Here, \emph{Alice} at table 1 turns ready to order a meal. Finally, Fig. 1e combines Fig. 1c and 1d. 







\subsection{A more involved step}
\begin{figure}[ht]
    \centering
    \includegraphics[scale=.25]{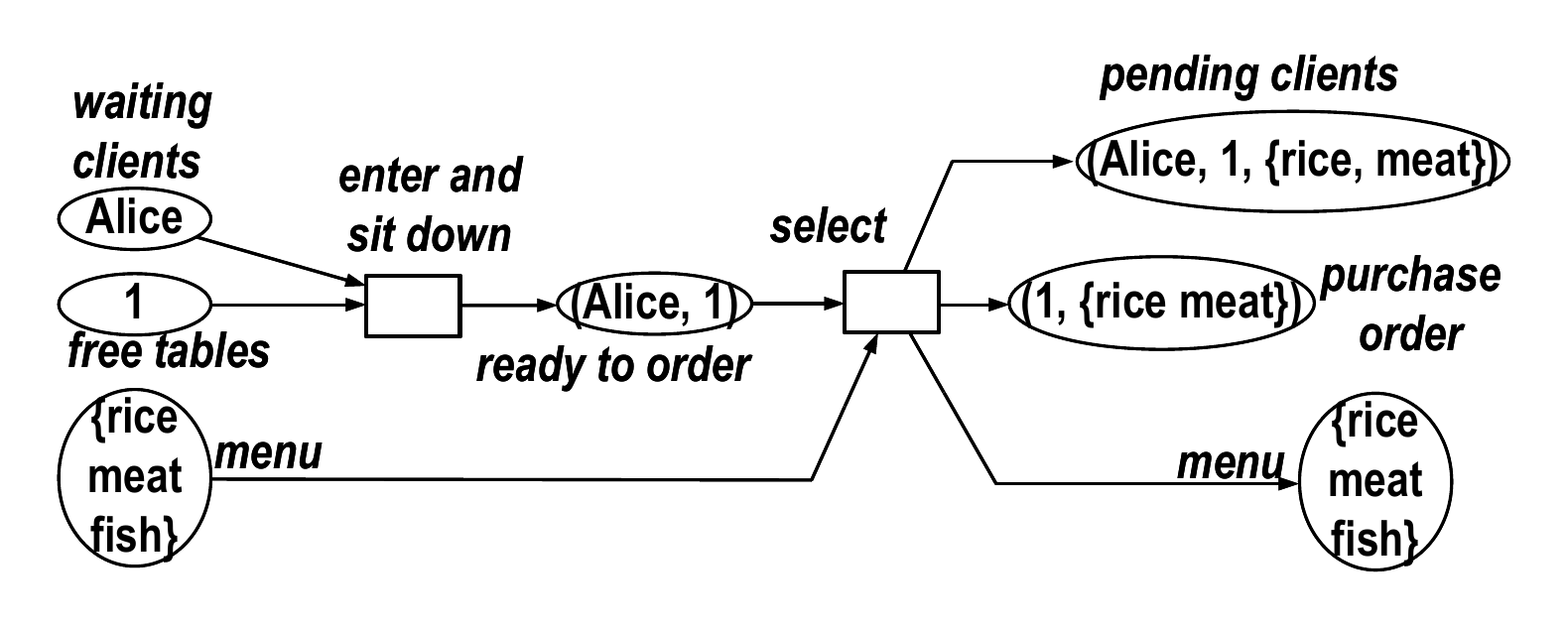}
    \caption{A more involved step}
    \label{fig:02}
\end{figure}

The step of Fig. 1e is now followed by another, more involved step. In this step, the client \emph{Alice} selects her meal (Fig. \ref{fig:02}). Here, the predicate menu represents the restaurant’s menu card. The menu card contains rice, meat, and fish. In our example, the predicate menu does not apply to the single dishes, but to the entire set of dishes. The step select makes use of this set, without updating it. \emph{Alice} selects the subset {rice, meat} from the menu and thus generates a purchase order for the kitchen. Coincidentally, \emph{Alice} sits pending at table 1.

\section{Conclusion}
This contribution emphasizes two aspects: firstly, that the role of modeling should be given more attention. Secondly, that modeling in informatics should be given a solid epistemic footing. We suggest heterogeneous algebraic structures, i.e. the mathematical basis of first order logic, as a most adequate such basis. 

As a fundamentally new concept, we suggest a dynamization of heterogeneous structures. Similar ideas have been applied for \emph{abstract state machines} previously denoted as \emph{evolving algebras} \cite{Gurevich_00}. From an axiomatic point of view, this choice is most thrifty: it only assumes the axioms of first order logic. \cite{meyer2025programmingreallysimplemathematics} emphasizes the advantages of such a starting point. Additionally, the close link to first order logic promises technically elegant and expressive means to deal with properties of digital systems.

This contribution describes the smooth path from classical, heterogeneous structures, to elementary steps of a dynamization. This is just a starting point; much more general constructs exploit the concept of predicates and a symbolic version of all this. Details can be found in \cite{Fettke_Reisig_24}.   

Summing up, we claim that \Heraklit is not just one out of many modeling infrastructures. With its carefully chosen footing it may be the base for many specializations, and a general theoretical base for modeling in informatics.

%
%
%

\bibliographystyle{splncs04}
\bibliography{main}





\end{document}